\documentclass[fleqn,twoside]{article}
   \usepackage{espcrc2}
   \usepackage{graphicx}
   \usepackage{amssymb}
   \usepackage{epstopdf}

\usepackage[figuresright]{rotating}

\def\teq#1{$\, #1\,$}                           
             \font\sevenrm=cmr7

          \font\sixrm=cmr6 

\def\today{\ifcase\month\or 
  January\or February\or March\or April\or May\or June\or 
  July\or August\or September\or October\or November\or 
  December\fi 
  \space\number\day, \number\year} 
\def\mathrm#1{{\hbox{\sevenrm #1}}} 
\def\aa{{Astron. Astrophys.}}

\def\apj{ApJ}
\def\apjsupp{ApJ Supp.}

\def\app{Astroparticle Phys.}                   
\def\apss{Astr. Space Sci.}                     
\def\grl{Geophys. Res. Lett.}                       

\def\mnras{{M.N.R.A.S.}}
\def\prl{Phys. Rev. Lett.}                      
\def\rmp{Rev. Mod. Phys.}                       
\def\ssr{Space Sci. Rev.}                       
\newcommand{\vol}[2]{$\,$\rm #1\rm , #2.}                 
\def\gamsk{\gamma_1}
\def\erg{\varepsilon_\gamma}
\def\dover#1#2{\hbox{${{\displaystyle#1 \vphantom{(} }\over{
   \displaystyle #2 \vphantom{(} }}$}}
\def\thetascatt{\theta_{\hbox{\sevenrm scatt}}}
\def\thetaBone{\Theta_{\hbox{\sevenrm Bn1}}}

{\catcode`\@=11 
  \gdef\SchlangeUnter#1#2{\lower2pt\vbox{\baselineskip 0pt\lineskip0pt 
  \ialign{$\m@th#1\hfil##\hfil$\crcr#2\crcr\sim\crcr}}}} 
\def\gtrsim{\mathrel{\mathpalette\SchlangeUnter>}} 
\def\lesssim{\mathrel{\mathpalette\SchlangeUnter<}} 

\newcommand{\AmS}{{\protect\the\textfont2
  A\kern-.1667em\lower.5ex\hbox{M}\kern-.125emS}}
  
\title{Diffusive Shock Acceleration of High Energy Cosmic Rays}

\author{Matthew G. Baring\address[MGB]{Department of Physics and Astronomy, MS-108\\
Rice University, P. O. Box 1892,\\
Houston, TX 77251-1892, USA}%
        \thanks{Email: {\it baring@rice.edu}}}

\begin{document}

\newcommand{\figureoutpdf}[5]{ \vspace{#1truein}
   \hspace{-0.0truein}
   \centerline{ \includegraphics[width=#3truein]{#2}}
   \hspace{-0.2truein}
   \vspace{#4truein} \caption{#5} }

\begin{abstract}
The process of diffusive acceleration of charged particles in shocked
plasmas is widely invoked in astrophysics to account for the ubiquitous
presence of signatures of non-thermal relativistic electrons and ions in
the universe.  A key characteristic of this statistical energization
mechanism is the absence of a momentum scale; astrophysical systems
generally only impose scales at the injection (low energy) and loss
(high energy) ends of the particle spectrum.  The existence of structure
in the cosmic ray spectrum (the "knee") at around 3000 TeV has promoted
contentions that there are at least two origins for cosmic rays, a
galactic one supplying those up to the knee, and even beyond, and perhaps an
extragalactic one that can explain even the ultra-high energy cosmic
rays (UHECRs) seen at 1-300 EeV.  Accounting for the UHECRs with
familiar astrophysical sites of acceleration has historically proven
difficult due to the need to assume high magnetic fields in order to
reduce the shortest diffusive acceleration timescale, the ion
gyroperiod, to meaningful values.  Yet active galaxies and gamma-ray
bursts remain strong and interesting candidate sources for UHECRs,
turning the theoretical focus to relativistic shocks.  This review
summarizes properties of diffusive shock acceleration that are salient
to the issue of UHECR generation.  These include spectral indices,
acceleration efficencies and timescales, as functions of the shock speed 
and mean field orientation, and also the nature of the field turbulence.   
The interpretation of these characteristics in the context of gamma-ray 
burst models for the production of  UHECRs is also examined.
\vspace{1pc}
\end{abstract}

\maketitle

\section{INTRODUCTION}

The origin and nature of very high energy
and ultra-high energy cosmic rays (UHECRs) continues to
pique the interest of the physics and astrophysics communities.  
This topicality is driven by the lack of resolution of the apparent 
incompatibility of the data taken by the AGASA and Fly's Eye 
experiments  at energies greater than around \teq{6\times 10^{19}}eV.
Moreover, both these experiments and the Yakutsk initiative claim 
detections \cite{AGASA94,FlysEye95,NW00} of cosmic ray events 
above the  so-called Greisen-Zatsepin-Kuzmin (GZK) cutoff, 
predicted by Greisen \cite{Greisen66} and Zatzepin
and Kuzmin \cite{ZK66}.  This observational characteristic can place
stringent constraints on both the seed for production of such
energetic particles, and the sites in the Universe for their generation.

The UHECR paradigm is also underpinned
by the existence of two competing scenarios for their creation:
(i) the more conventional bottom-up models, where the particles
are accelerated from much lower energies up to the extreme values
indicated by the air shower array data, and (ii) the generally newer and
more exotic ``top-down'' possibilities that invoke creation and
decay of various entities in particle physics that cascade
down in energy to the GZK domain.  Top-down scenarios have emerged
as competitors to the more traditional bottom-up concepts since the
latter have difficulty achieving energies in the \teq{10^{20}}eV range
in most known astrophysical objects.

This review focuses on the properties of diffusive acceleration
at relativistic shocks that are relevant to the production of UHECRs 
in astrophysical settings.  These characteristics are not widely known 
outside the cosmic ray acceleration field.
The viability of any putative source of UHECRs is contingent on its
ability to generate particles of requisite energies in the available
time, and with sufficient abundances and appropriate metallicity.    
Such quantities are sensitive to assumed parameters of shocks, 
such as field obliquity, level of turbulence, the nature of scattering, 
and the role of accelerated particles in shaping the shock 
hydrodynamics.  Subtle (and even not so obscure) 
implications of acceleration properties are often not addressed 
in bottom-up models for UHECRs that are offered for peer scrutiny, 
being left to specialist discussions within the acceleration community.   

Here, the goal is to highlight some key properties of relativistic shock 
acceleration that are pertinent to various UHECR scenarios.  To
prepare the way, a few standard results from acceleration theory at
non-relativistic shocks are briefly reviewed in Section~\ref{sec:NR}, 
to facilitate comparison with the relativistic domain.    This is
the most-studied aspect of shock acceleration theory, with a number
of very instructive reviews \cite{Drury83,BE87,JE91,MD01} in the
literature.  Moreover, it is the eminently testable domain, affording
observational diagnostics via {\it in situ} spacecraft measurements
\cite{EMP90,BOEF97} of accelerated populations at shocks in the 
heliosphere, and also numerous particle production and radiation 
models of supernova remnants (see \cite{Baring00} for a review),
the principal contender for the sources of galactic cosmic rays.  
Section~\ref{sec:rel} then turns to the intricacies of relativistic shocks,
focusing on several characteristics that distinguish them from their
non-relativistic counterparts.  A discussion of the impact of these
properties on the gamma-ray burst (GRB) paradigm for UHECR
production is offered in Section~\ref{sec:impact}.

\section{NON-RELATIVISTIC SHOCK ACCELERATION THEORY}
 \label{sec:NR}

The theory of diffusive acceleration of particles at non-relativistic
shocks (those with upstream fluid speeds \teq{u_1\ll c} in the shock
rest frame) has been thoroughly investigated by both
analytic and simulation techniques.  The process, also know as
Fermi acceleration, proceeds by particles being transported back and forth
across a shock between colliding fluid flows.  The kinematics of 
momentum deflections then yields a net energy gain, and the process
can continue for a large number of shock crossings to generate high
energy cosmic rays before they are lost due to convection, upstream
escape, or even cooling.  Much is understood about this process,
though the key outstanding issue is how electrons and ions are heated
in the shock layer, and subsequently injected into the diffusive acceleration
process.  It is instructive to review three basic characteristics of 
\teq{u_1\ll c} shocks to set the scene for the relativistic considerations
of Section~\ref{sec:rel}.

\subsection{Canonical Index}

Non-relativistic shocks generate particles with a power-law distribution
in momentum.  This is a consequence of high energy
particles (those with speeds \teq{v \gg u_1}) attaining isotropy in all
pertinent reference frames, so that the so-called {\it diffusion 
approximation} can be applied.  At such energies, the principal
transport equation describing the acceleration process, 
the diffusion-convection equation, can be solved analytically for
plane shocks \cite{BO78,JE91}, yielding 
the well-known result for the momentum distribution 
\begin{equation}
   \dover{dn}{dp} \, \propto\, p^{-\sigma} 
   \quad \hbox{with} \quad
   \sigma = \dover{r+2}{r-1} \ ,
 \label{eq:TPpowerlaw}
\end{equation}
where \teq{r=u_1/u_2} is the shock (velocity) compression ratio, 
\teq{p} is the momentum.  Eq.~(\ref{eq:TPpowerlaw}) is a 
steady-state, test-particle result.  In this limit, the spectral index, 
\teq{\sigma}, is independent of the shock speed, \teq{u_1}, the 
field obliquity, and any details of the scattering process as long as 
isotropy of highly super-thermal particles is maintained.  The 
canonical nature of this result is a driving force behind invocations 
of acceleration in astrophysics.  Note that a high Mach number,
non-relativistic shock has \teq{r\approx 4} and \teq{dn/dp\propto p^{-2}}.

\subsection{The Character of Oblique Shocks}

The result in Eq.~(\ref{eq:TPpowerlaw}) exhibits no information 
concerning the normalization of the power-law.  Such a property is 
controlled by the rate at which particles are injected from thermal
energies.  This injection rate is sensitive to the angle \teq{\thetaBone}
the mean upstream magnetic field makes to the shock normal.
In the downstream region, particles are swept away more efficiently
from the shock by the convective force of the flow when the
obliquity angle \teq{\thetaBone} is higher.  For non-relativistic shocks
of high Mach number, when \teq{\thetaBone\gtrsim 30^\circ}, the 
convection is so effective that injection of thermal particles
into the acceleration process is entirely suppressed \cite{BEJ94}.
The critical angle at which acceleration ceases increases
to around \teq{\thetaBone\sim 55^\circ} for hotter, low Mach number
shocks. These results apply if the diffusion is dominant along the
magnetic field, i.e. the ratio \teq{\kappa_{\perp}/\kappa_{\parallel}}
of spatial diffusion coefficients \teq{\kappa_{\parallel}} along
the field and \teq{\kappa_{\perp}} orthogonal to {\bf B} is much 
less than unity.  Here \teq{\kappa_{\parallel ,\perp} =\lambda_{\parallel ,\perp} v/3}
for scattering mean free paths \teq{\lambda_{\parallel ,\perp} (v)}, 
for particle speeds \teq{v}.  A concomitant effect
of obliquity is the dramatic increase in the acceleration rate
\cite{Jokipii87,Ostrow88} in highly oblique shocks, above the
\teq{\thetaBone =0^\circ} values
\cite{FJO74} that can be inferred from Eq.~(\ref{eq:timaccfit}) 
below. This increase can
be ascribed to the presence of electric fields in the shock layer that seed
shock-drift acceleration.

While the rapid reduction of the acceleration time in
oblique and quasi-perpendicular shocks is enticing, the 
inefficient acceleration is unattractive to cosmic ray production
models.  This trade-off is an unavoidable property of oblique
shocks \cite{EBJ95}.  Restricting
to \teq{\kappa_{\perp}\ll \kappa_{\parallel}} cases is not always
appropriate.  Increasing \teq{\kappa_{\perp}}, physically corresponding to
stronger field turbulence, tends to eliminate laminarity information
in the field and accordingly renders the shock more like a parallel,
\teq{\thetaBone\sim 0^\circ} one.  Consequently, increasing \teq{\kappa_{\perp}}
towards the Bohm limit of \teq{\kappa_{\perp}\sim \kappa_{\parallel}}
mutes the increases in acceleration rates, and returns injection efficiencies 
to levels commensurate with plane-parallel shocks \cite{EBJ95}.  
Modelers therefore must compromise: either opt for quasi-laminar 
oblique shocks that are faster but less efficient accelerators, or 
extremely turbulent shocks of arbitrary obliquity that generate cosmic
rays efficiently but at standard rates, i.e. the inverse gyrofrequency
corresponding to a specific particle energy.

\subsection{Nonlinear Modifications}

The aforementioned features apply to circumstances where the
accelerated particles do not modify the shock hydrodynamics,
i.e. are dynamically unimportant and act as test particles.
Yet, when acceleration is extremely efficient, a sizable
fraction of the total energy budget emerges as high energy cosmic
rays, an inevitable occurrence in an \teq{r \approx 4} shock if the power-law
in Eq.~(\ref{eq:TPpowerlaw}) extends to high enough energies.
The pressure of these particles decelerates the upstream flow,
which in turn provides feedback to the distribution
of accelerated ions and electrons, and therefore the fraction of energy 
going into these particles.  This nonlinearity has been thoroughly
explored in the literature 
\cite{DV81,Drury83,Eichler84,EE84,EBJ96,Berezhko96,Malkov97,Blasi02} 
and is a critical
characteristic of efficient shocks.  The quintessential example is
the Earth's bow shock immersed in the solar wind; it affords a nice 
data comparison between experiment and acceleration 
theory \cite{EMP90}.

As a result of the energy conservation that regulates the acceleration
and the energy apportionment between thermal ions and high energy
cosmic rays, there are two distinctive features of nonlinear,
cosmic ray modified shocks: (i) a distribution that deviates from pure
power-law nature, exhibiting a characteristic upward concavity due to higher energy
particles sampling larger effective compression ratios, since their
diffusive mean free paths are longer, and (ii) the thermal particles
are somewhat cooler \cite{BE99} than for test-particle shocks, since the subshock
is weakened and energy removed from the thermal ions and electrons.
These phenomena can be probed to a certain extent by examining
isolated, radiating systems such as supernova remnants, and at present, 
there are at best modest indications to support these theoretical predictions.

\section{DISTINGUISHING PROPERTIES OF RELATIVISTIC SHOCKS}
 \label{sec:rel}
 
Diffusive acceleration at relativistic shocks is far
less studied than that for non-relativistic flows, yet it is
a most applicable process for UHECR generation, and
may occur in extreme objects such as pulsar
winds, hot spots in radio galaxies, jets in active galactic 
nuclei and microquasars, and GRBs. 
Early work on relativistic shocks was mostly analytical in the
test-particle approximation (e.g., \cite{Peacock81,KS87a,HD88,KW88}),
although the analytical work of \cite{SK87,BK91} explored
nonlinear, cosmic ray modified shocks.  
Complementary Monte Carlo techniques have been employed for 
relativistic shocks by a number of authors, including test-particle 
analyses by \cite{KS87b,EJR90} for parallel,
steady-state shocks, and extensions to include oblique magnetic fields by
\cite{Ostrow91,BH92,BO98}.

A key characteristic that distinguishes relativistic shocks from their
non-relativistic counterparts is their inherent anisotropy
due to rapid convection of particles through and
away downstream of the shock, since particle speeds \teq{v} are never
much greater than the downstream flow speed \teq{u_2\sim c/3}.
Accordingly, the diffusion approximation, the
starting point for virtually all analytic descriptions of shock
acceleration when $u_1\ll c$, cannot be invoked since it requires nearly
isotropic distribution functions.  Hence analytic approaches prove more
difficult when \teq{\gamsk \gg 1}, though advances in special cases such as
the limit of extremely small angle scattering ({\it pitch angle diffusion})
are possible \cite{KS87a,Kirketal00}.  Let us explore
some of the distinctive properties of particle acceleration at relativistic shocks.

\subsection{Non-Universality of the Spectrum}

The most attractive feature of non-relativistic shock acceleration 
theory is that the distribution of accelerated particles is
scale-independent, i.e. a power-law, as in Eq.~(\ref{eq:TPpowerlaw}), 
with an index \teq{\sigma} 
that depends only on the velocity compression ratio \teq{r=u_1/u_2}, 
i.e. hydrodynamic quantities.  This elegant result does not carry over 
to relativistic shocks because of their strong plasma anisotropy.  As a 
consequence, while power-laws are indeed created, 
the index \teq{\sigma} becomes a function of the 
flow speed, the field obliquity, and the nature of the scattering, all
of which intimately control the degree of particle anisotropy.

In the specific case of parallel, ultrarelativistic
shocks, the analytic work of Kirk et al. \cite{Kirketal00} demonstrated
that as \teq{\Gamma_1\to\infty}, the spectral index \teq{\sigma}
asymptotically approached a constant, \teq{\sigma\to 2.23}, a value
realized once \teq{\Gamma_1\gtrsim 10}.  This enticing result,
which has been confirmed by Monte Carlo simulations
\cite{BO98,Baring99,AGKG01,ED02}, has been
referred to sometimes as an indication of the universality of the index
in relativistic shocks.  In this subsection, it is illustrated that the 
asymptotic index of 2.23 is
indeed not canonical, but rather a special case corresponding to
compression ratios of \teq{r=3} and the particular assumption of 
small scattering (pitch angle diffusion), specifically for incremental 
changes \teq{\thetascatt} in a particle's momentum with angle
\teq{\thetascatt \ll 1/\Gamma_1}.

First, the spectral index of the power-law distribution is a declining
function of the Lorentz factor for a fixed compression ratio, a characteristic 
evident in \cite{KS87a,BH91,Kirketal00} for the case of pitch angle 
scattering, and a property that extends to large angle scattering 
\cite{EJR90,Baring99}.  Faster shocks generate flatter distributions 
if \teq{r} is held constant, a consequence of the increased kinematic 
energization occurring at relativistic shocks.  Note that
imposing a specific equation of state such as the J\"uttner-Synge one
renders \teq{r} a function of \teq{\Gamma_1} so that this monotonicity
property can disappear, as evinced in Fig.~2 of Kirk et al. \cite{Kirketal00}.
Table~\ref{tab:tab1} lists indices \teq{\sigma}
of the \teq{dn/dp} distribution obtained from the Monte Carlo
simulation technique of Ellison et al. \cite{EJR90}.  These results, obtained
specifically in the limit of pitch angle diffusion, illustrate the flattening
as \teq{\Gamma_1} increases.

\vspace{-10pt}
\begin{table}[htb]
\caption{Spectral Indices $\sigma$ for Pitch Angle Diffusion
                at Relativistic, Plane-Parallel Shocks.}
  \label{tab:tab1}
 \renewcommand{\tabcolsep}{4.5pt} 
  \renewcommand{\arraystretch}{1.2} 
\begin{tabular}{lccccc}
\hline
$\Gamma_1\beta_1\; ^a$ & $r=2$ & $r=2.5$ & $r=3$ & $r=3.5$ & $r=4$ \\[2pt]
\hline
30    &  3.24 &  2.57 &  2.23  & 1.99  & 1.86  \\
10    &  3.28 &  2.59  & 2.23  & 2.00 &  1.87  \\
3      &  3.33  & 2.64  & 2.26  & 2.02  & 1.86  \\
2      &  3.38  & 2.67  & 2.28  & 2.03  & 1.88  \\
1      &  3.48  & 2.72  &  2.31 &  2.05  & 1.89  \\
0.3   &  3.90  & 2.95  &  2.42 &  2.15  & 1.96  \\
0.1   &  3.96  & 2.98  &  2.46 &  2.17  & 1.98  \\
0.03$^b$  &  3.98  & 2.99  &  2.49  &  2.19  & 1.99  \\
\hline
\end{tabular}\\[3pt]
The spectral indices are from the Monte Carlo simulation of
\cite{EJR90}, and the accuracy of their determination is typically
of the order of \teq{\pm 0.02}. Notes: (a) Here \teq{\beta_1=u_1/c} and
\teq{\Gamma_1} are the dimensionless velocity and Lorentz factor of
the upstream flow in the shock rest frame, respectively.  (b) 
This non-relativistic limit approximately reproduces the well-known
 \teq{\sigma = (r+2)/(r-1)} result in Eq.~(\ref{eq:TPpowerlaw}).
\vspace{-10pt}
\end{table}

The choice of the canonical compression ratio \teq{r=3} is a well-known
result for a relativistic, purely hydrodynamic shock possessing
an ultrarelativistic equation of state \cite{BM76}.  However, one can
envisage situations where the magnetic field becomes dynamically
important.  The classic example is the termination shock for
the Crab pulsar wind, where Kennel \& Coroniti \cite{KC84} observed that
strong fields can weaken magnetohydrodynamic shocks
considerably.  In an interesting generalization of this, Double et al. 
\cite{Double04} recently determined deviations from $r=3$ in 
ultrarelativistic shocks, in cases where pressure anisotropy is significant,
a characteristic that is expected to be common in relativistic shocks.
Such departures can either strengthen {\it or} weaken the shock
depending on the nature of the pressure anisotropy, which must be 
a significant function of the shock obliquity, i.e., \teq{\thetaBone}.  
Hence, we anticipate that \teq{\sigma} will be a function  
\teq{\thetaBone}, an issue visited again in this Section.

\begin{figure}[htb]
\figureoutpdf{-0.5}{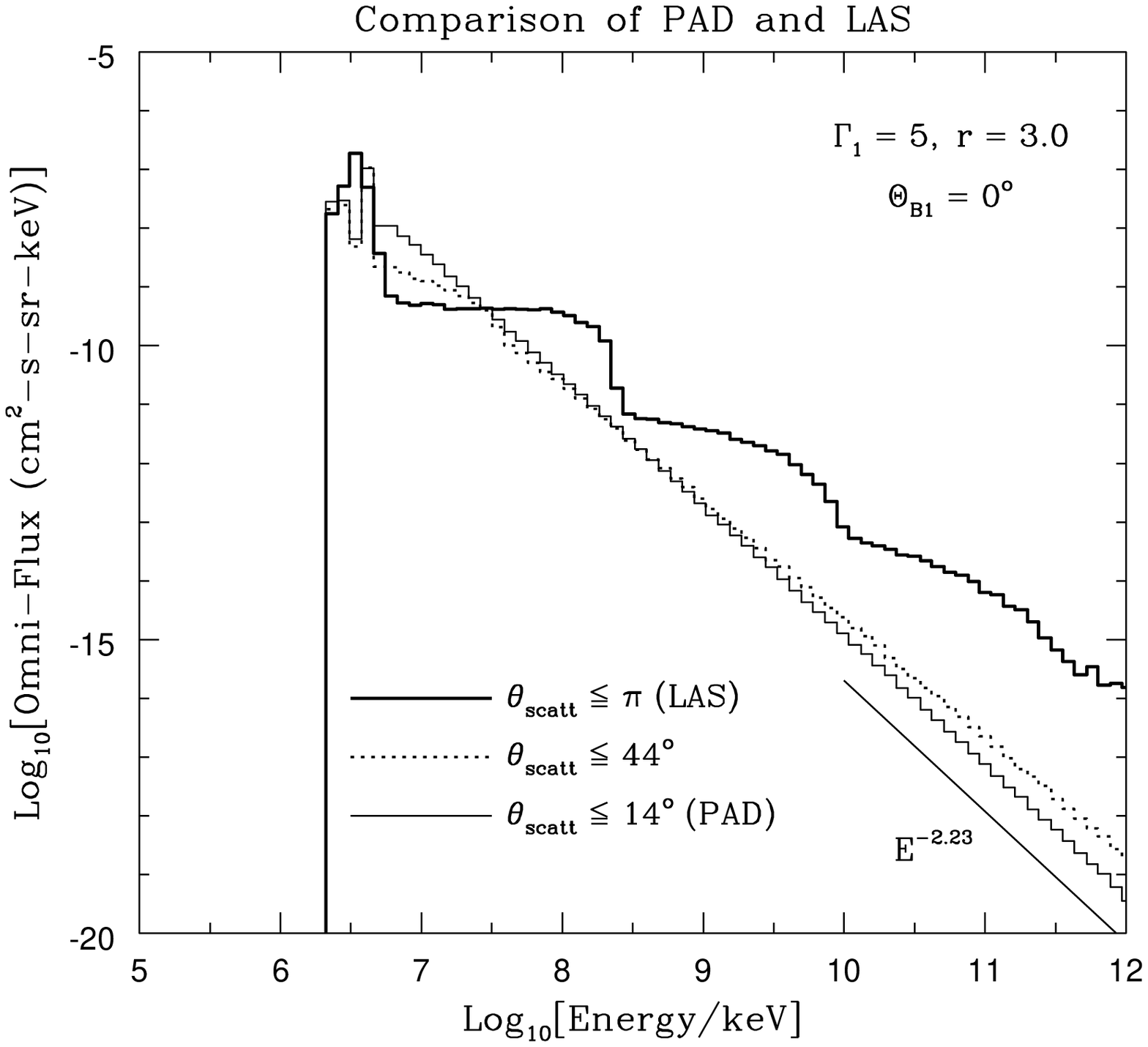}{3.5}{-0.9}{
Particle distributions from a parallel (\teq{\thetaBone=0}) 
relativistic shock of \teq{r=3} and Lorentz factor
\teq{\Gamma_1=5}, obtained from a Monte Carlo simulation
\cite{EJR90,Baring99}.  Scattering is modeled by randomly
deflecting particle momenta by an angle
\teq{\thetascatt} within a cone whose axis coincides with the momentum
prior to scattering.  Distributions are depicted for three cases,
\teq{\thetascatt \leq 14^\circ}, corresponding to pitch angle diffusion
(PAD), large angle scattering (LAS: \teq{\thetascatt\leq\pi\gg
1/\Gamma_1}), and an intermediate case (dotted histogram).
 \label{fig:spectra} } 
\end{figure}

More novel is the fact that the slope of the nonthermal particle
distribution depends on the nature of the scattering, a feature evident
in the works of Refs. \cite{EJR90,BO98,Baring99}.  The asymptotic,
ultrarelativistic index of 2.23 is realized only in the mathematical limit
of pitch angle diffusion (PAD), where the particle momentum is
stochastically deflected on arbitrarily small angular (and therefore temporal)
scales.  In practice, PAD results when the scattering angle 
\teq{\thetascatt} is inferior to the Lorentz cone angle \teq{1/\Gamma_1}
in the upstream region.  In such cases, particles diffuse in the
region upstream of the shock only until their angle to the shock normal
exceeds around  \teq{1/\Gamma_1}.  Then they are rapidly swept
to the downstream side of the shock.  The energy gain per shock
crossing cycle is then of the order of a factor of two,
simply derived from relativistic kinematics \cite{GA99,Baring99}.  The
spectrum, depicted in Fig.~\ref{fig:spectra}, is then slightly steeper 
than the \teq{p^{-2}} result for a strong, non-relativistic shock, 
due to the balance between particle energization and loss by 
convection downstream.

The results in Fig.~\ref{fig:spectra} are from the Monte Carlo 
simulation of acceleration at relativistic shocks developed by 
Ellison et al. \cite{EJR90}, who demonstrated that for
large angle scattering (LAS, with \teq{\thetascatt\sim\pi}) the
spectrum is highly structured and much flatter than 
\teq{E^{-2}}.  Such a case is exhibited in the Figure.  The
structure is kinematic in origin, where large angle deflections
lead to distribution of fractional energy gains between unity
and \teq{\Gamma_1^2}.  Gains like this are kinematically 
analogous to the
energization of photons by relativistic electrons in inverse
Compton scattering.  Each structured bump or spectral
segment in Fig.~\ref{fig:spectra} corresponds to an increment
in the number of shock crossings, successively from \teq{1\to3\to
5\to 7} etc., as illustrated by Baring \cite{Baring99},
that eventually smooth out 
to asympotically approach an index of \teq{\sigma\sim 1.5}.
Clearly, such highly-structured distributions have not been
inferred from radiation emission in any astrophysical objects.

An intermediate case is also depicted in Fig.~\ref{fig:spectra},
with \teq{\thetascatt\sim 4/\Gamma_1}.  The spectrum is smooth,
like the PAD case, but the index is lower than 2.23.  Astrophysically,
there is no reason to exclude such cases.  Moreover, from the
plasma point of view, magnetic turbulence could easily be
sufficient to effect scatterings on the order of these angles, a
contention that becomes even more salient for ultrarelativistic shocks
with \teq{\Gamma_1\gg 10}.  Note that the \teq{\Gamma_1=5}
results depicted here are entirely representative of the nature 
of such ultrarelativistic cases.  Clearly
a range of indices can be supported when \teq{\thetascatt}
is chosen to be of the order of \teq{1/\Gamma_1}, and the scattering
corresponds to the transition between the PAD and LAS limits.
It is anticipated that various astrophysical systems will encompass 
a range of scattering properties.  Accordingly, the continuous and
monotonically decreasing behavior of \teq{\sigma} with 
\teq{\thetascatt}, as indicated in the
exposition of \cite{ED04}, highlights the non-universality of
the distribution index in relativistic shocks.

Categorizing the scattering as either PAD or LAS is a useful
division, but is not a complete description of diffusive transport
in shocks.  Another characteristic, the diffusion of particles
across mean field lines, becomes a critical element in the 
discussion of oblique or perpendicular shocks.   As mentioned 
above, when the upstream angle \teq{\thetaBone} of the field 
to the shock normal is significant,  diffusion of particles in the 
downstream region struggle to compete 
with convective losses, and transport back upstream of 
the shock layer becomes inefficient.   In
non-relativistic shocks, this effect was explored by Baring et al.
\cite{BEJ94}, who observed that the losses controlled the
injection efficiency of nonthermal particles, so that when 
\teq{\thetaBone\gtrsim 30^\circ}, thermal particles of speed
\teq{v\gtrsim u_1} fail to return to the shock after one crossing 
to the downstream side and the Fermi acceleration process
is quenched.  Accordingly, for \teq{\thetaBone < 30^\circ} in such
regimes, while power-law superthermal distributions are
exhibited, their normalization is a strongly declining function
of \teq{\thetaBone} when transport across field lines is
suppressed.    

\begin{figure}[htb]
\figureoutpdf{-0.5}{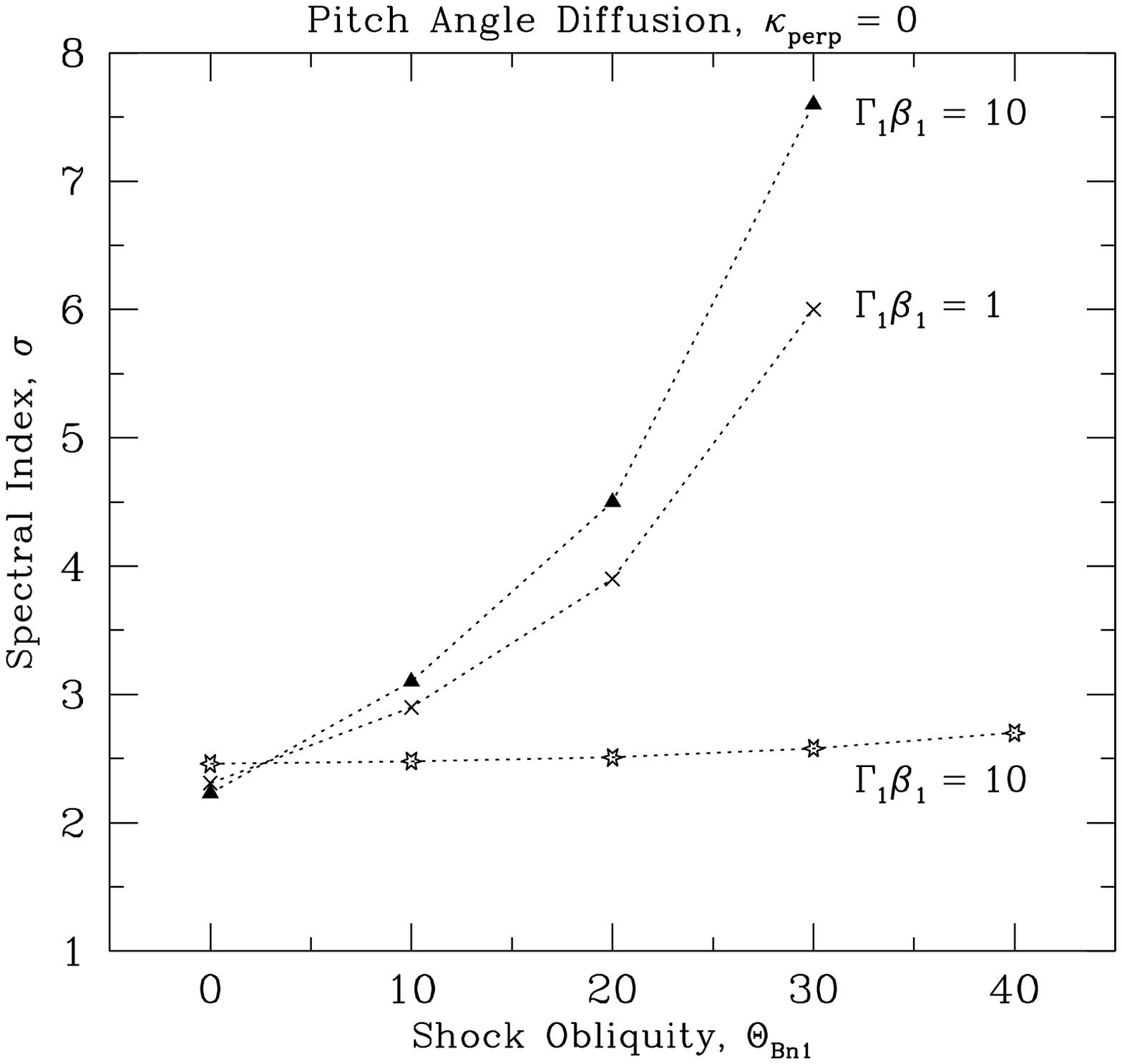}{3.5}{-0.9}{
Particle distribution indices \teq{\sigma} from oblique (\teq{\thetaBone >0}) 
relativistic shocks of \teq{r=3} and different Lorentz factors
\teq{\Gamma_1}, obtained from a Monte Carlo simulation
\cite{EJR90,Baring99} in the limit of pitch angle diffusion.  
Results are depicted for the case of
zero diffusive transport perpendicular to the mean field, i.e.,
\teq{\kappa_{\perp}=0} for the perpendicular component of the
spatial diffusion coefficient \teq{\kappa}.  The index is insensitive
to \teq{\thetaBone} for non-relativistic shocks, but rapidly increases
with obliquity for relativistic ones, underlining their inherent inefficiency. 
 \label{fig:obq_index} } 
\end{figure}

This phenomenon is manifested in a somewhat different manner
in relativistic shocks.  When \teq{u_1\sim c}, the \teq{v\gtrsim u_1} 
criterion for
dramatic, if not catastrophic, convective losses is satisfied
for {\it all} particle speeds, not just slightly suprathermal
ones.  Hence such losses can be expected to be pervasive
for all non-thermal energies.  Increased losses must diminish
the nonthermal population, and since the loss rate is purely a function of
particle speed \cite{Peacock81,JE91}, which is effectively pinned
at \teq{v\approx c},  and \teq{u_2}, the net effect is to increase the 
spectral index while retaining power-law character.  This property is
illustrated in Fig.~\ref{fig:obq_index}, where the simulation output
was acquired in the absence of cross field diffusion (i.e.,
\teq{\kappa_{\perp}=0}).
Increasing \teq{\thetaBone} results in a rapid rise in \teq{\sigma}
corresponding to a suppression of acceleration.  
Essentially, for \teq{\thetaBone\gtrsim 25^\circ}, acceleration
is virtually non-existent for \teq{\Gamma_1\beta_1\gtrsim 1}.

Turbulent plasmas in shock environs generally will not admit a
\teq{\kappa_{\perp}=0} assumption.  Strong turbulence will drive 
the system towards the so-called {\it Bohm-diffusion} limit, 
where diffusion coefficients are similar parallel and perpendicular 
to the field, i.e. \teq{\kappa_{\perp}\sim \kappa_{\parallel}}, and 
transport is effectively isotropic.  Efficient transport across field 
lines can to a significant extent circumvent convective losses, 
returning the particles to the shock from the downstream region, 
and accordingly flattening the power-law distribution.  It is anticipated
that transport near the Bohm limit would be essential to generate
\teq{\sigma\lesssim 3}, i.e. indices meaningful for UHECR acceleration
paradigms.  This is borne out in \cite{BO98} and the recent work of 
Ellison \& Double \cite{ED04}, who obtained \teq{\sigma\approx 2.34} for a
\teq{\Gamma_1=10}, \teq{\thetaBone=60^\circ} shock in the extreme case
of the Bohm limit.  The Monte Carlo simulation results of 
\cite{BO98,ED04} also exhibited the expected monotonic decrease 
in \teq{\sigma} with the increase in \teq{\kappa_{\perp}/ \kappa_{\parallel}}.

\subsection{Acceleration Times}

Having explored spectral issues germane to the UHECR problem,
we now turn our attention to the maximum energy issue.  This
is essentially determined by the rate at which particles are accelerated,
so diffusive acceleration times become the focal point.   Various authors
have researched this subject for relativistic shocks 
\cite{EJR90,BO96,Bednarz00,AGKG01,Baring02,MQ03}.  In particular,
\cite{EJR90} found that
for large angle scattering, the acceleration time for a
\teq{\Gamma_1\lesssim 5} shock was only marginally shorter than that
expected from classical non-relativistic shock theory.  The simulations of
Bednarz \& Ostrowski \cite{BO96,Bednarz00} revealed similar modest
reductions for both LAS and PAD.   Observe that in spite of substantial
energy gains per shock crossing, typically on the order of \teq{\Gamma_1^2},
the particles then spend considerable time diffusing downstream, a time
coupled to their inverse gyrofrequency.

Recently, Baring \cite{Baring02} 
computed acceleration times in the limit of pitch angle diffusion
using the simulation of \cite{EJR90}, specifically for application to
jets in blazars.  It was found that extrapolation of simulations into the relativistic
regime revealed a hard lower bound on the total acceleration time
\teq{\tau_{\rm acc}} {\it as measured in the shock rest frame}.  
The time \teq{\tau_{\rm acc}} monotonically decreases (for
ultrarelativistic particles) to this limit as \teq{\Gamma_1} increases
to infinity, yet proximity is achieved for \teq{\Gamma_1\gtrsim 10}.
If \teq{\nu_g} represents the energy-dependent gyrofrequency of an
ultrarelativistic electron or ion, then the velocity dependence of the
acceleration times in plane-parallel shocks can
be approximated (to around 1--3\% accuracy) by the empirical fit 
\begin{eqnarray}
  \tau_{\rm acc} &\approx& \Biggl(\dover{1}{4}
         - \dover{0.18}{\Gamma_1\beta_1}
         + \dover{1}{\Gamma_1^2\beta_1^2} 
         + \dover{0.22}{1+\Gamma_1\beta_1} \Biggr)
           \tau_{\hbox{\sixrm NR}}\nonumber \\[-5.5pt]  
 \label{eq:timaccfit}\\[-5.5pt]
   && \tau_{\hbox{\sixrm NR}}\;\equiv\; 
          \tau_{\hbox{\sixrm NR}} (\beta_1=1)
          \; =\; \dover{f}{\nu_g}\;\; .\nonumber
\end{eqnarray}
Here \teq{\tau_{\hbox{\sixrm NR}}(\beta_1=1)} is the extrapolation 
of the well-known acceleration time formula \cite{FJO74,Jokipii87}
for non-relativistic, parallel shocks to flow speeds \teq{c}.  The times are 
for a velocity compression ratio of \teq{u_1/u_2=3},
and the coefficient \teq{f} describes details of the differences in
diffusion between the upstream and downstream regions, and is of the
order of unity and independent of \teq{\Gamma_1}.  Note that
when \teq{\beta_1\ll 1}, the familiar non-relativistic result emerges:
\teq{\tau_{\rm acc}=\tau_{\hbox{\sixrm NR}} /\beta_1^2}.  Introducing 
shock obliquity can speed up the acceleration, as in 
non-relativistic cases, but at the price of dramatically steepening
the distribution.

The bound arises due to the insensitivity of the downstream flow speed
and diffusion in the downstream region to the upstream \teq{\Gamma_1}.
Downstream diffusion yields the dominant
contribution to \teq{\tau_{\rm acc}}, with upstream particles requiring
only small deflections (accomplished in short times: e.g., \cite{Bednarz00}) 
from the shock normal in order to return downstream.  This 
automatically implies a hard lower bound to \teq{\tau_{\rm acc}} as
\teq{\Gamma_1\to\infty}, since the downstream speed saturates at
\teq{c/3}.   Effectively, particles can never be
accelerated at rates much faster than their gyrofrequency.  The limit
translates to a {\it comparable limit in the upstream fluid frame}, which
is often the observer's reference perspective, for example the interstellar 
medium surrounding a jet.  This property 
follows from the connection between Lorentz
transformations of times and energies, with the proper time of the
particle being a Lorentz invariant.  

Hence, models of acceleration at relativistic shocks do not incur
any increases to the energization rate other than the enhancement
of the field by a single Lorentz boost.  Maximum energies are then
only explicitly weakly dependent on shock speeds.  This implies that
sites for cosmic ray acceleration generally must invoke higher
environmental magnetic fields to effect higher maximum particle
energies.

\section{IMPACT ON UHECR PARADIGMS}
 \label{sec:impact}
 
As an illustrative and topical case, the focus here is on the scenario that
gamma-ray bursts are sites for the generation of ultra-high energy 
cosmic rays.  The discussion will emphasize acceleration issues, as
opposed to source population considerations.

\subsection{Gamma-Ray Burst Applications}

The paradigm that gamma-ray bursts (GRBs) are responsible for
UHECR production \cite{MU95,Waxman95,Vietri95}, has been
quite topical over the last decade.  Bursts are sufficiently 
energetic to amply satisfy cosmic ray energy budgets if their space 
density is not too sparse.  Consequently the redshift distribution of GRBs,
not well-known at present but soon to be refined by the SWIFT mission, 
is an important constraint \cite{SS02} on the ability of bursts to serve 
as sites for UHECR generation.  Given the level of community interest, 
it is salient to assess the aforementioned acceleration results in the 
context of gamma-ray bursts.  

First, the discussions above indicate that the maximum energy of 
cosmic rays from bursts is not increased by subtle relativistic effects,
and can be approximately estimated using standard non-relativistic
shock theory with modification as per Eq.~(\ref{eq:timaccfit}).  It
can be quickly deduced, by comparing the inverse gyrofrequency 
with subsecond burst variability timescales, that UHECRs 
can be generated in GRBs if the magnetic fields inside the burst
are of the order of $B \sim 10^5-10^7$G inside the emission region.
This is not dissimilar from field estimates (e.g., \cite{Meszaros02,BB04}) 
obtained by  synchrotron radiation modeling of their prompt 
gamma-ray emission in the MeV band, so approximate consistency 
is achieved.  It is not yet understood how such large fields arise in 
GRB shocks, though ideas of field amplification at shocks 
\cite{ML99,LB00} have recently become prominent.

The non-universality of the power-law index and its sensitivity to 
obliquity and the anisotropy of turbulent transport immediately 
indicate that GRB spectra should possess diverse indices.  This 
should be manifested in the energy range above the 100 keV -- 1 MeV 
peak of emission, and is in fact so in data taken from the 
EGRET experiment on the Compton Gamma-Ray Observatory (CGRO), 
where the half dozen or so bursts seen at high energies
have a broad range of spectral indices \cite{Dingus95}, namely
\teq{\alpha \sim 2-3.7} for \teq{dn/d\erg\propto \erg^{-\alpha}}.
This result suffers from limited statistics due to (i) the nature of
bursts, and (ii) to EGRET's field of view being more limited than that
for BATSE, the principal GRB experiment on CGRO.  The
upcoming GLAST mission will provide a more refined determination
of the distribution of burst indices above 30 MeV after its launch in 2007.

At the same time, in order to emit the radiation detected, GRBs
must have underlying electron distributions that are relatively flat.
By extension, since the electron and cosmic ray ion distributions
trace each other in theories of shock acceleration where 
the diffusive mean free path is dependent only on rigidity, then
the ion index must lie in the range \teq{\sigma \sim 3-6}.  This
is a spectral constraint commensurate with that imposed by the
observed $E^{-3}$ UHECR 
distribution \cite{Watson00,NW00}.  As the UHECR spectrum
results from a convolution of a host of sources, modulo unknown
propagation effects, one expects that
the flatter distributions will dominate, so there is a satisfying
consistency between UHECR ion and GRB photon spectra.  
Since GRB shocks are believed to be ultrarelativistic, the
acceleration results discussed above (as exemplified in 
Fig.~\ref{fig:obq_index}) indicate that
strong cross field diffusion (i.e., \teq{\kappa_{\perp}\sim 
\kappa_{\parallel}}) will be necessary in the majority
of bursts, if their shocks are oblique, as is highly likely.  This
property is required to provide a cosmic ray distribution 
at least as flat as the observations, and indeed efficiently 
generate cosmic rays in sufficient numbers.  Note that the
same would apply to jets in active galaxies such as blazars,
as an alternative source of UHECRs.

An acceleration issue not addressed above concerns the shape 
of the particle distributions at thermal and slightly suprathermal
energies.  This is essentially an injection or dissipational heating
issue that is readily probed for electrons by the spectrum of 
prompt GRB emission.  Baring \& Braby \cite{BB04} pursued
a program of spectral fitting of GRB emission using a linear
combination of thermal and non-thermal electron populations.
These fits
demanded that the preponderance of electrons that are responsible for the
prompt emission reside in an intrinsically non-thermal population,
strongly contrasting particle distributions obtained from acceleration
simulations. This result implies a conflict for acceleration
scenarios where the non-thermal electrons are drawn directly from a
thermal gas (the virtually ubiquitous case), unless radiative efficiencies 
only become significant at
highly superthermal energies.  Another potential caveat is that strong
radiative self-absorption could be acting, in which case the GRB 
spectral probe is not sampling the thermal electrons.  Considerable
work is needed to resolve this issue to ascertain whether an
acceleration paradigm can be truly consistent with the GRB 
emission that is seen.

\section{OUTLOOK}

The UHECR field is clearly anticipating the next generation of
observational data from the Auger experiment, just around the corner.
In the meantime, theorists will continue to develop their models
and hone their understanding.  In terms of acceleration theory, 
at least two key developments can be expected in the coming 
years.  

First, with the advent of rapidly increasing computational
capability, reliance on analytic and Monte Carlo techniques is
no longer essential, and these acceleration approaches will be 
supplanted in part by modeling by fully 3D plasma codes, namely
particle-in-cell (PIC) simulations.  Such simulations have 
generated interesting results in the last decade, but have been
hampered \cite{JKG93,JJB98} by restricted dimensionality imposed
by CPU memory and speed limitations.  In the last 2-3 years, 
results from 3D codes have achieved increasing visibility, 
particularly for modeling relativistic shocks \cite{Silva03,Nishikawa03}.  
Their key impediment is having to treat
widely disparate inertial scales associated with the proton to
electron mass ratio.  Their virtue is the accurate modeling of
particle heating, coherent acceleration effects and field amplification
in the shock layer.  One can expect many interesting developments
and results from PIC codes in the coming decade.

Monte Carlo techniques will still continue to be powerful tools,
due to their capability of handling large dynamic ranges in
spatial and momentum scales.  Thus they are ideally suited to
nonlinear acceleration scenarios, which can impact the interpretation of
UHECR ion composition studies, since nonlinear non-relativistic systems
are well-known to preferentially accelerate ions with higher
mass to charge ratios \cite{EDM97}.
The application of the Monte Carlo approach to nonlinear,
relativistic shocks has so far been very limited \cite{ED02}, and
this is territory ripe for investigation.  It is anticipated that nonlinear
effects will be more subtle for the \teq{u_1\sim c} domain, since
particle distributions are sensitive to the shock speed, obliquity
and the type of scattering that operates.  In addition, the shock 
hydrodynamics are dependent on the plasma anisotropy, and this
is inextricably connected to \teq{u_1}, \teq{\thetaBone} and
\teq{\thetascatt} also.  Hence, research in the near future should
focus on elucidating the interplay between these ingredients.  Such
simulations also need to explore ways to address injection and
heating of thermal species in a more consistent manner, including
the influence of electric potentials in the shock layer.

\end{document}